\documentclass[a4paper,10pt]{article}

\usepackage{moreverb}

\usepackage{amsmath,amssymb,enumerate}
\usepackage{cite,graphics,graphicx}
\usepackage{url}
\usepackage{balance}
\usepackage{listings}
\newcommand\BibTeX{{\rmfamily B\kern-.05em \textsc{i\kern-.025em b}\kern-.08em
T\kern-.1667em\lower.7ex\hbox{E}\kern-.125emX}}

\usepackage[T1]{fontenc}
\usepackage{authblk}

\begin{document}


\title{Desktop to Cloud Migration of Scientific Computing Experiments}

\author[1]{Satish Narayana Srirama}
\author[1]{Pelle Jakovits}
\author[2]{Vladislav~Ivani{\v s}t{\v s}ev}
\affil[1]{Mobile \& Cloud Lab, Institute of Computer Science, University of Tartu, J. Liivi 2, Tartu, Estonia}
\affil[2]{Institute of Chemistry, University of Tartu, Ravila Str. 14A, Tartu, Estonia}
\renewcommand\Authands{ and }



\affil[ ]{\textit {\{srirama, jakovits, vladislav.ivanistsev\}@ut.ee}}

\maketitle

\begin{abstract}

Scientific computing applications usually need huge amounts of computational power. The cloud provides interesting high-performance computing solutions, with its promise of virtually infinite resources on demand. However, migrating scientific computing problems to clouds and the re-creation of software environment on the vendor-supplied OS and cloud instances is often a laborious task. It is also assumed that the scientist who is performing the experiments has significant knowledge of computer science, cloud computing and the migration procedure, which is often not true. Considering these obstacles, we have designed a tool suite that migrates the complete software environment directly to the cloud. The developed desktop-to-cloud-migration (D2CM) tool supports transformation and migration of virtual machine images, reusable deployment description and life-cycle management for applications to be hosted on Amazon Cloud or compatible infrastructure such as Eucalyptus. The paper also presents an electrochemical case study and computational experiments targeted at designing modern supercapacitors. These experiments have extensively used the tool in drawing domain specific results. Detailed analysis of the case showed that D2CM tool not only simplifies the migration procedure for the scientists, but also helps them in optimizing the calculations and compute clusters, by providing them a new dimension -- cost-to-value of computational experiments.

\end{abstract}


\section{Introduction}

Scientific computing is a field of study that applies computer science to solve typical scientific problems. It is usually associated with large-scale computer modeling and simulation and often requires a large amount of computer resources. Cloud computing \cite{ArmbrustETAL:AboveCloud.09} suits well in solving these scientific computing problems, with its promise of provisioning virtually infinite resources.  Cloud computing is a style of computing in which, typically, resources scalable on demand are provided "as a service (aaS)" over the Internet to users who do not need to have knowledge of, expertise in, or control over the cloud infrastructure that supports them. The provisioning of the cloud services can be at the Infrastructural (IaaS), Platform (PaaS) or Software (SaaS) levels.


Cloud computing is intertwined with its three main features: elasticity, utility computing, and virtualization. A cloud platform dynamically provisions, configures, reconfigures, and deprovisions computing resources on demand and in real-time. The elasticity of a framework can be defined as its ability to adjust according to the varying loads of requests or requirements it has to support. Elasticity and the cloud's promise of provisioning virtually infinite resources make it ideal for resource-intensive scientific computing tasks. Utility computing model forwards the idea of computing as a utility, where consumers pay based on how many resources they use. This allows the scientists to have no upfront commitment costs and allows to estimate the cost of experiments before actually proceeding with them. Last but not the least of the features is the virtualization \cite{Barham:2003:XAV:1165389.945462} technique, which is used to separate the hardware and software and allows running several independent virtual servers on a single physical device. 

Significant research~\cite{srirama2011scalability, mapscale, buyya.2009} has been performed in migrating scientific computing applications to the cloud. A number of science clouds \cite{HelixNebula.13, SriramaETAL:SciCloud.10, FutureGrid.13} have been established in recent years, in which scientists have run many scientific simulations and applications and measured their performance to evaluate the applicability of doing science on the cloud. However, the idea of running scientific computing applications on the cloud has not been well received by all quarters of the scientific computing community, as the performance of cloud still lags behind when compared to grids or clusters. 

This is because communication and other types of latency added by the virtualization technology are major hindrances when executing scientific computing applications on the cloud~\cite{srirama2011scalability, SPE:SPE995}. The cloud computing community is trying to address these issues, and several solutions have already been proposed \cite{SPE:SPE995, DBLP:conf/synasc/LiG10}, in recent years. The support of high-performance computing (HPC) with Cluster Compute Instances by the public clouds like Amazon EC2 is also a positive step in this direction.

While the research of migrating scientific computing experiments to the cloud shows great potential, most often it is assumed that the users who port their applications to the cloud have significant knowledge of computer science, cloud computing, and the migration procedures. However, from our observations this is not the case with non-computer scientists. For example, scientists who are running their computational modeling and simulation tasks on grids and clusters are unaware how the environment is configured, how the queues work on clusters and how their jobs get executed. All they are interested in is that they submit a job to some queue and after some time they can collect the results. Having this assumption in mind, we have designed Desktop-to-Cloud-Migration (D2CM)~\cite{Srirama:MCLab.D2CM, srirama2013direct} tool suite that helps scientists to migrate their experiments to the cloud. The idea is to migrate the complete software environment, in which the scientists have set up their experiments, directly to the cloud.

D2CM tool integrates several software libraries to facilitate VM image transformation, migration, and management of computing experiments running in the cloud. To be able to leverage the cloud resources, we specifically concentrate on migrating distributed applications, such as applications which use the Message Passing Interface (MPI) to parallelize the execution of the computations, but it can be used for any distributed solution or standalone application. The tool also supports monitoring the health of the cluster while the application is running in the cloud. 

Once the prototype of the D2CM tool was ready, we used it to migrate several of our scientific computing experiments to the cloud. This paper addresses our experience with migrating an electrochemistry case study to the cloud. Grid-base Projector-Augmented Wave Method (GPAW)~\cite{GPAW2} electronic structure calculator was used for the computing experiments. GPAW is a popular distributed program suit which in conjunction with the Atomic Simulation Environment (ASE) \cite{ase} provides optimized tool-set for electronic structure calculations. The GPAW program was used to study the capacitance of an interface between a metal electrode and a room-temperature ionic liquid. The computing experiments we performed with the aim to improve supercapacitors' and batteries' performance \cite{ivanistsev2012} are explained briefly along with their field-specific necessity. In the process, we also discuss the experiments performed by the electrochemists to show how they could scale their trials by successfully exploiting the cloud resources and characteristics, using the D2CM tool.

The rest of the paper is organized as follows. Section 2 describes the cloud migration procedure in detail along with the developed tool. Section 3 briefly explains the considered electrochemical case study while the conducted computing experiments and some of the observed results are summarized in section 4. Section 5 provides an overview of the related work. Section 6 concludes the paper with future research directions.

\section{Desktop to Cloud Migration}
A major hurdle in migrating applications to an IaaS cloud is the re-creation of a software environment on the vendor-supported OS and cloud instances. The D2CM tool seeks to mitigate potential problems of this often laborious task by taking the entire environment as-is, provided it is within a virtual environment. The D2CM tool supports non-interactive applications which operate on a finite input data set and it provides three main facilities: 

\begin{itemize}
\item{Programmatical transformation of user-supplied virtual machine images to the target cloud image format. Currently, it has support for migrating VirtualBox disk images (VDI) to Amazon Machine Image (AMI) format and to Eucalyptus~\cite{Euca:site} compatible XEN \cite{Barham:2003:XAV:1165389.945462} images.}
\item{Life-cycle management of distributed applications, hosted on Amazon Elastic Compute Cloud (EC2) \cite{Amazon:EC2} platform or compatible infrastructure, such as Eucalyptus.}
\item{Creating deployment templates to specify how exactly the application is to be deployed on the cloud.}
\end{itemize}


\subsection{Virtual image transformation}
\label{vmtransform}
EC2's and Eucalyptus' underlying machine virtualization is XEN (Note: Eucalyptus also supports KVM), a widely used open-source technology which supports full and para-virtualization. While Amazon provides a range of already configured images, it also supports user-supplied images, provided they function with a Xen compatible kernel. The kernel must have pv-ops with XSAVE hypercall disabled \cite{amazonkernels}. An image may either be configured to boot with Amazon supplied kernels, known as AKIs, or with a kernel packaged within the image itself. The later option is used by D2CM, which uses Xen's PVGRUB boot process. In the kernel update section of the transformation process, the D2CM tool installs a new kernel, either 32 or 64 bit depending on the detected image type, and modifies GRUB configuration to boot with the new kernel by default. In the current implementation, the tool only supports the transformation of a single disk partition. If an image has more than one partition, only the primary one can be migrated to EC2.

D2CM uses the Libguestfs library \cite{libguestfs} to inspect and manipulate virtual machine images. Using libguestfs is highly preferred to mounting the images within the user's filesystem as directly mounting the image requires the host OS to support the image's filesystem(s) and requires root privileges. Libguestfs' API includes functions to extract partitions from images and to insert SSH key files, which are required during the image migration process. Libvirt \cite{libvirt} is an abstraction library for interacting with desktop hypervisors. The D2CM tool uses libvirt to access the currently supported hypervisor, VirtualBox, which is an open source hypervisor that runs on Windows, Linux, Macintosh, and Solaris hosts and supports a large number of guest operating systems.   In the future, adding support for other popular hypervisors such as KVM, Xen, and VMWare should be greatly eased by the use of libvirt. The currently supported IaaS platforms, EC2 and Eucalyptus, provide the same Amazon Web Services (AWS) API for managing machine instances. The web service is wrapped by several libraries which provide programming language specific bindings. As the D2CM tool is developed using Python, we used the Boto library which is the Python binding for AWS.

In addition to working with a compliant kernel, an instance created from this image must also be able to mount devices in the manner supplied by the EC2 virtual machine. EC2 instances have a default device mapping that usually does not match the fstab entries within the user-supplied image. EC2 instances come in a variety of types. 
Some instances do not provide swap devices, while, most instances provide additional storage devices that the user may want to utilize. To conform to this new virtual hardware environment, D2CM modifies the fstab appropriately. 

\subsection{Virtual image migration}

Before the transformation process, D2CM tool goes through the following steps to actually transform and migrate a user-supplied image. 

\begin{itemize}

\item{\textbf{Specifying cloud credentials:}
The first step to application migration requires the user to supply the tool with valid AWS credentials so that the tool can access cloud services. For Amazon EC2 and Eucalyptus, these credentials include user id, access key, secret key, user certificate file location and private key file location.}



\item{\textbf{Image registration and migration:}
The user then registers a desktop image with the tool by choosing a name, an image file located in the local machine and the default cloud for this image. The user can then select to initiate the migration process, by choosing the target cloud, its availability region, and bucket name in the cloud file system, where the virtual machine image will be stored.} 


\item{\textbf{Image transformation:}
The tool proceeds with the image transformation steps, making any changes to a separate image file, to preserve the integrity of the source image. (See figure~\ref{fig:transformation})}

\begin{figure}
\centering
\includegraphics[width=0.70\textwidth]{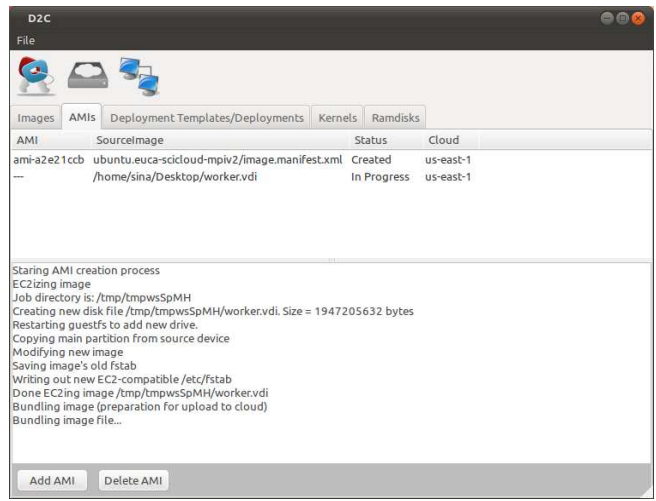}
\caption{Image transformation view in the D2CM tool}
\label{fig:transformation}
\end{figure}

\item{\textbf{Image upload and registration:}
The tool uploads the transformed image to Amazon's Simple Storage Service (S3) \cite{amazon:s3} or to Walrus in the case of Eucalyptus. The tool then registers the image with the EC2/Eucalyptus, declaring default settings such as suitable system architecture, default kernel, and ramdisk, default instance type, etc. All images are registered by default as private, only authorizing read and boot access to the account which was used to upload and register the image.

The duration of the entire transformation and upload procedure can vary greatly depending on the size of the image and internet connection speed. At this point, the user's image is ready to run on the cloud. This process must be repeated for each unique desktop image required for the distributed application.}

\end{itemize}

\subsection{Defining deployment template}
\label{deploymenttemplate}

\begin{figure}
\centering
\includegraphics[width=0.70\textwidth]{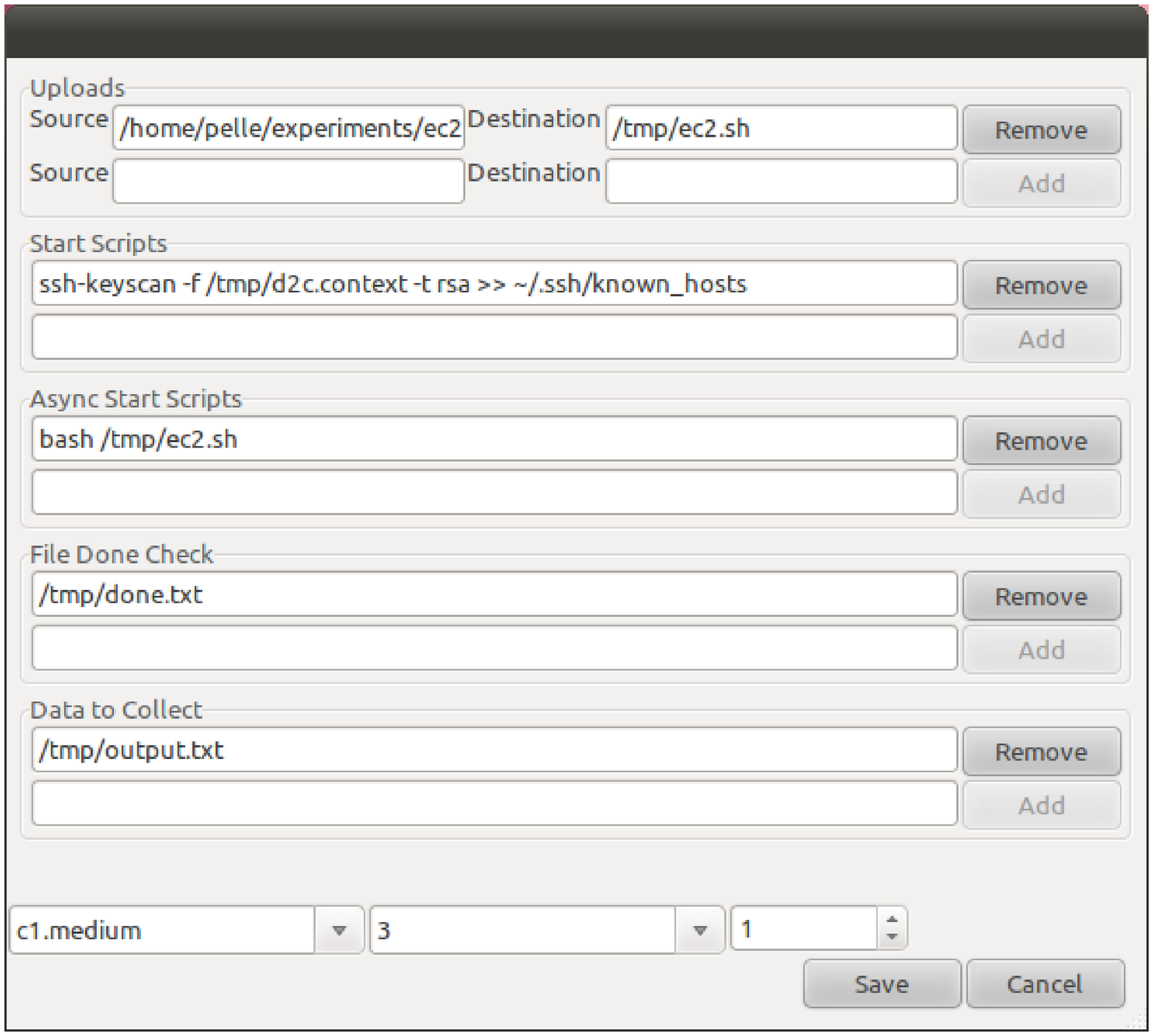}
\caption{Defining a role in the D2CM tool}
\label{fig:role}
\end{figure}

Next step after migrating the virtual machines is defining a deployment template that describes how the virtual machines are divided into roles and how these roles are executed on the cloud. A deployment consists of one or more roles and each role is defined by a name, association to a previously migrated cloud image, instance count, instance type and life-cycle scripting. The number of roles is not limited, and there may be a many-to-one relationship between roles and a unique cloud image. The instance count defines the number of virtual machines that will be allocated upon deployment start-up for a role. Roles include the following optional life-cycle commands that the deployment manager invokes at the appropriate stage:

\begin{itemize}
\item{Input files that should be uploaded from the local machine to the virtual machines.}
\item{Start-up command(s) that are executed when the virtual machine boots.}
\item{Initiation commands that specify which command starts the asynchronous execution of the applications (in our case, computing experiments) inside the virtual machines.}
\item{Ending condition for checking when the computing experiment has been completed.}
\item{Download actions that specify which files are downloaded once the computing experiment is done.}
\end{itemize}

\begin{figure}
\centering
\includegraphics[width=0.70\textwidth]{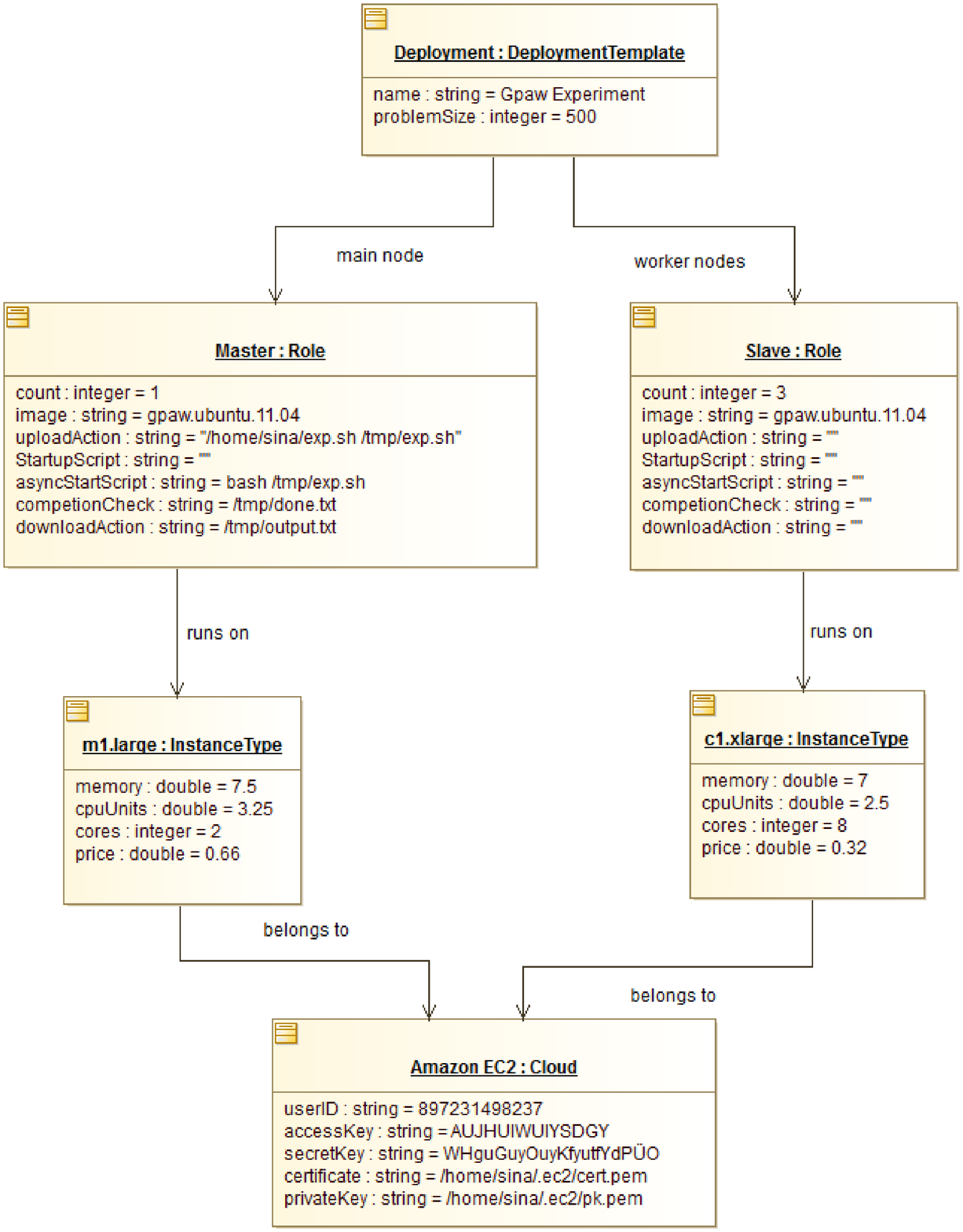}
\caption{An example deployment template model}
\label{fig:deploymentTemplate}
\end{figure}

Figure~\ref{fig:deploymentTemplate} illustrates a deployment template model that we used for executing scientific computing experiments in Amazon EC2. Additionally, figure~\ref{fig:role} shows the actual user interface for defining the deployment roles in the D2CM tool. 

\subsection{Executing a deployment on the cloud}

D2CM supports non-interactive applications which operate on a finite input data set. The D2CM tool does not directly support interactive applications which may prompt for some kind of instructions or responses from the user. For such applications, the user can directly log into the cloud instances from their local machine via ssh. 

The defined deployment template describes a specific type of experiments and to execute a single experiment on the cloud the user has to create a deployment instance from the deployments and initiate it's execution. The started deployment and its life-cycle proceed through the following phases in the D2CM tool.

\subsubsection{Reservation} 
Upon initiation, the tool iterates through deployment roles and reserves instances matching user specifications. If the request cannot be completely fulfilled by the cloud provider, any successful allocation is released and the process ends. AWS begins charging the user's account, the moment instances are reserved. Prior to the actual reservation, the tool prompts the user for confirmation, displaying the per hour rate in USD as determined by the deployment's instance types and count, to confirm that the user is prepared to pay this price. 

\subsubsection{Contextualization}
Processes that operate cooperatively require some discovery mechanism and the tool facilitates this through contextualization step. When all instances within a deployment are assigned IP addresses, the tool makes sure every instance has access to all the IPs in the deployment, through configuration files. 

\subsubsection{Start-up}
Once the instances have been acquired, the tool executes any user-defined start actions. If a role contains multiple instances, the start actions will be run on each instance. Currently, the tool supplies three classes of start-up actions: synchronous, asynchronous, and upload. 

The upload action allows the user to install files on the instances which were not bundled with the image. These files may either correct minor image defects or supply computing experiment specific input. The synchronous action executes a specified BASH script. The tool serially executes all scripts for a given role and its instances. This functionality allows users to make modifications to deployment hosts that are guaranteed to finish before starting any main scripts, which are run using the asynchronous actions. The asynchronous action executes a script remotely on an instance, returning execution control immediately back to the tool. In contrast to the synchronous action, the user should assume that all asynchronous actions run in parallel. The D2CM tool logs on to the remote instance, executes these commands in a daemonizing (with $nohup$) wrapper and immediately ends the remote connection. This is the mechanism by which a user should initiate the main application process. An example asynchronous action is:

$mpirun~-np~16~python~matrix{\_}calc.py$

Once all asynchronous actions have been initiated, the tool changes the deployment state to completion monitoring.

\subsubsection{Completion monitoring}

The user must configure the D2CM tool to test for application-specific completion. The only option currently provided is the ability to monitor for the presence of a user specified file on an instance. The tool continually polls the instances of roles with the specified check. If no check is provided for a given role, that role is automatically assumed to be in the finished state. The program proceeds to finalization only when all the roles are in the finished state.

\subsubsection{Finalization}

The result of supported applications will be a collection of output files. These must be migrated to persistent storage as the cloud instances only provide ephemeral storage, i.e. storage that exists for a limited life-span of the instance. D2CM supports configurable downloading of the user specified files from the instances to the local storage. 

\subsubsection{Termination}
Upon retrieval of the required files, the deployment is shut down. D2CM terminates all instances associated with the deployment, ending any of the runtime charges from the cloud provider. 

\subsubsection{Recovery from failures}

D2CM stores the running state of experiments in a local database and in a case of network or machine failures, it can reconnect to the cloud infrastructure and restore the state of the experiment. The user can shut down the tool while an experiment is running and continue to monitor it later, which is necessary for long-running experiments. This feature is critical especially as the experiments are performed from the desktop.

\subsection{Results and re-execution of finished experiments}

\begin{figure}
\centering
\includegraphics[width=0.70\textwidth]{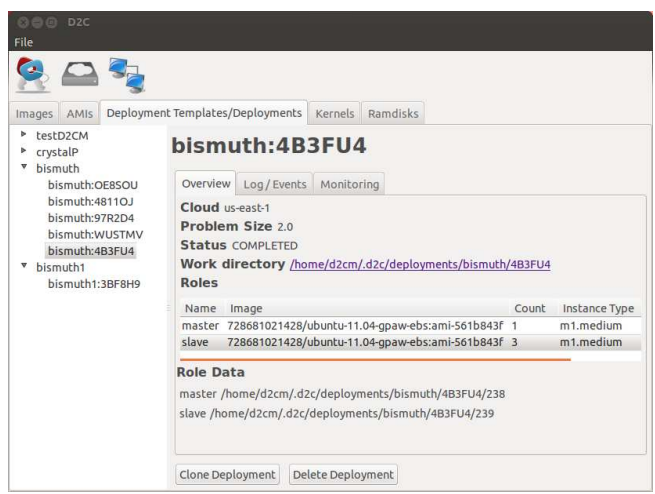}
\caption{A completed deployment shown in the D2CM tool}
\label{fig:completed}
\end{figure}

Once the experiment is completed, the tool provides a link to the results as shown on the figure \ref{fig:completed}. It points to a "`work directory"' on the local machine, where all the results (download actions specified in the deployment template) were downloaded, which can be accessed by the user to verify the results of the experiment. 

D2CM tool keeps track of all the previously run experiments and the user can re-execute any of them by simply cloning the deployment instance, changing some parameters and deploying them. The user can also run multiple experiments simultaneously and the tool can keep track of all of them. This provides convenient means to execute a chain of experiments which only differ by some parameter values. It also helps to scale up the deployment infrastructure between experiments without having to create new deployment descriptions from scratch.

\section{Migration Case Study}
\label{casestudy}

The goal of this research is to provide effective tools for conducting computational experiments on a cloud. To validate the approach, we ported a computational electrochemistry case study to the cloud. 

Electrochemistry is an actively studied branch of chemistry, and some of its most well known applied solutions are batteries, fuel cells, supercapacitors and solar panels. Several Noble prizes were awarded during the past decade for discoveries concerning electrochemistry. One of them was shared between Walter Kohn for his development of the ``density-functional theory (DFT)'' and John A. Pople for his development of ``computational methods in quantum chemistry'' \cite{nobel1998}. Another was shared between Martin Karplus, Michael Levitt, and Arieh Warshel ``for the development of multiscale models for complex chemical systems'' \cite{nobel2013}. The use of DFT-based computational methods to understand electrochemical processes by examining molecular scale models is continuously gaining popularity within the electrochemistry community \cite{Calle-vallejo2012,Anderson2012}.

\begin{figure}
\centering
\includegraphics[width=0.65\textwidth]{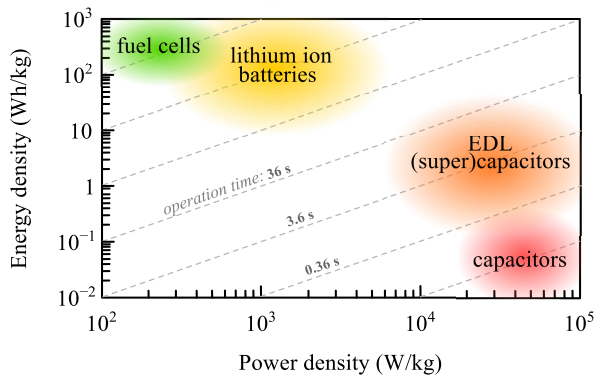}
\caption{Comparison of lithium battery and supercapacitor}
\label{fig:LitVsSuperCap}
\end{figure}

In the case study, we investigated room temperature ionic liquids \cite{Fedorov2014} by the means of computational electrochemistry methods to find a better electrolyte for supercapacitors \cite{Conway1999}. Ionic liquids are a novel type of electrolytes with a unique combination of properties that can be used to improve the performance of batteries, fuel cells, supercapacitors and solar panels \cite{Fedorov2014}. Supercapacitors have several advantages over other energy storage and transformation devices: they have impressive charge--discharge time (less than 10 seconds); virtually unlimited number of life-cycles (over one million); high specific power (10 kW/kg); and excellent operating temperature range ($-20$ to $65^\circ\rm{C})$ \cite{lewandowski2004}. However, when compared to common lithium-ion batteries, their cost per $\mathrm{W\times h}$ is ten times larger and the amount of energy they can store is ten times lower. The performance of capacitors and batteries is compared in figure \ref{fig:LitVsSuperCap}, representing energy density vs. power density chart for various energy-storage devices. ``Right and higher'' is better in the chart; data for lithium-ion batteries are in the upper left corner, and data for supercapacitors are on the right.  

The energy density ($E$) of supercapacitors can be improved in accordance with the following equation:

\begin{equation} 
\label{eq:energy}
\begin{centering}
E = \frac{1}{2}C\Delta U^2,
\end{centering}
\end{equation}

\noindent where $C$ is capacitance and $\Delta U$ is voltage.

Currently, the voltage is almost the same for both lithium battery and supercapacitor, but with ionic liquids it can be increased for the later by two times from 3 Volts to almost 6 Volts \cite{lewandowski2004}. Moreover, it is also possible to increase the capacitance. Herewith, the scientific challenge is to reach the best possible characteristics at the laboratory conditions. In this process, our computer modeling may hint at the direction for the experimental research. The engineering challenge is to improve characteristics of modern supercapacitors and simultaneously reduce their price by designing them in such a way, that they would be commercially competitive with modern batteries and thus would become available at the market. A successful research and development of novel prototypes have been conducted at the University of Tartu \cite{lust2003,kurig2011}, laying the ground for the world's leading high-performance supercapacitor producer Skeleton Technologies \cite{tartutech}. We contribute to this success by performing computational modeling of ionic liquids at electrode surface as described below.

\subsection{Computational modeling}

\begin{figure}
\centering
\includegraphics[width=0.65\textwidth]{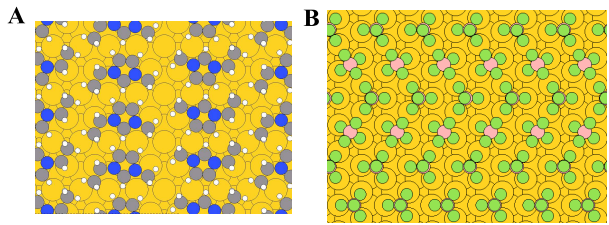}
\caption{A. Molecular model for a cationic (1-ethyl-3-methylimidazolium) layer at negative Au(111) surface; B. Molecular model for an anionic (tetraflouroborate) layer at positive Au(111) surface}
\label{fig:structureModel1}
\end{figure}
\begin{figure}
\centering
\includegraphics[width=0.65\textwidth]{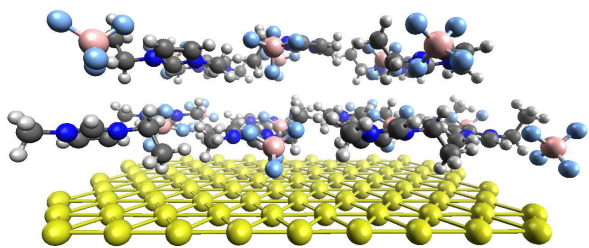}
\caption{Molecular model of an interface between neutral Au(111) surface and 1-ethyl-3-methylimidazolium tetraflouroborate}
\label{fig:structureModel2}
\end{figure}

We started our computational study with a fairly simple model of Au(111) \textbar~$\textnormal{EMImBF}_4$ interface by representing the interface as a \textit{gold slab} (electrode surface) and a \textit{single adlayer of ions} (figure \ref{fig:structureModel1}) \cite{ivanistsev2012}. Figure \ref{fig:structureModel2} shows a model with a thick (2~nm) overlayer of an ionic liquid. This model is consistent with experimental results, which were obtained with the help of scanning tunneling microscope, electrochemical impedance spectroscopy and cyclic voltamperometry \cite{pan2006,gnahm2010}. These results suggest that at positive electrode potentials a structured layer of anions is formed; at negative potentials cations form an ordered layer structure, and in between, both anions and cations are present at the surface. The latter case requires a complex model, shown in figure \ref{fig:structureModel2}, and was omitted in the presented study.

We investigated the two preceding cases separately by performing DFT calculations with the help of GPAW tool \cite{GPAW,GPAW2}, which is described in the following subsection in more detail. Starting from a given geometry, the GPAW calculator finds optimal geometry with the lowest energy and evaluates the electron density, which is then used to evaluate absolute electrode potential. 

The model used (figure \ref{fig:structureModel1}) assumes domination of only one type of ions (characterized by coverage -- a variable number of ions per surface area unit) which completely compensate the charge at the electrode surface. According to the DFT formalism, the charge density is redistributed during the calculation procedure so that the electric field sets up between the surface and ions. In this way we get a simple molecular-scale capacitor, which to a certain degree mimics the real interface between an electrode and an ionic liquid. According to the Molecular Dynamics simulations results, at high positive and negative electrode potentials a dense structured monolayer of ions at electrode surface coexists with apparently non-structured ionic liquid further from the electrode \cite{Ivanistsev2014,Ivanistsev2015}. This observation hints to a possible reference point between the models used in Molecular Dynamics simulations as well as in our DFT calculations and the real interface. We reasonably assume that at other potentials our models provide a good guess of interfacial characteristics. For qualitative comparison, in section \ref{results}, we focus on the interfacial free energy and Bader-type charges of ions.

According to the formula \ref{eq:energy}, a higher energy density can be achieved by increasing the capacitance and voltage. To do so, it is required to understand how these two parameters depend on structure and composition of the interface between ionic liquid and electrode. This goal can be achieved either experimentally or theoretically through computational modeling. Within the experimental approach, very complex systems are studied predominantly at macroscopic level, while the computational modeling operates with very simple systems represented by molecular-scale models. Therefore, the challenge of obtaining reliable results from computational experiments is correlated with balancing between the model complexity and the amount of available resources. A sophisticated model obviously accounts for more effects and thus is more realistic, but is also much more resource demanding. 

To obtain most realistic results, which would be comparable to the experimental electrochemical results, one needs high-CPU resources, a migration tool, quantum chemical calculator and a reliable model. In our study, these are the cloud resources, D2CM, GPAW program suite, and the model justified above, respectively.

\subsection{GPAW program}
Computational routines within GPAW program are based on rather novel Grid-based Projector-Augmented Wave method \cite{GPAW,GPAW2}. The code is released under the GNU Public License (GPL) and has several distinguished features, which are necessary for electronic structure calculations of large electrochemical systems via parallel computing. 
Firstly, there are at least two ways of representing the wave functions: 1) using lcao-mode, which gives less accurate results, but computationally is more efficient and provides a good guess for optimized geometry; 2) using ``finite difference'' or plane-wave mode, which gives more accurate results, but is highly demanding on computational resources.
Secondly, the parallelization with MPI can be done over the $k$-points, bands, spin in spin-polarized calculations, and using real-space domain decomposition. 
Finally, the time-dependent DFT implemented in GPAW program enables calculation of dielectric function, optical and X-ray adsorption spectra. In addition to vibrational spectrum, these in principle allow a direct comparison to experimentally measured data. With the main focus on the D2CM migration process, in this study we proceed with optimization of several systems, without performing any sophisticated time-dependent DFT calculations. 

Setup for a single computational experiment is defined via the atomic simulation environment (ASE) python syntax \cite{ase}, which is used to define atomic positions, calculator routines used, optimization procedures, input-output data etc. The following script is an example of one of our computational experiments:

\begin{lstlisting}
  #Define geometry of metal slab
  slab = fcc111('Au', size=(2,2,3),
                orthogonal=True)
  slab.center(axis=2, vacuum=12)
  d=0.8
  a=asin(1/sqrt(3))+pi
  #Define geometry of BF4 anion
  BF4 = Atoms([Atom('B', ( 0, 0, 0)),
               Atom('F', ( d, d, d)),
               Atom('F', (-d,-d, d)),
               Atom('F', (-d, d,-d)),
               Atom('F', ( d,-d,-d))])
  #Change orientation of BF4
  BF4.rotate('y',pi/4,center=(0,0,0))
  BF4.rotate('x',a,   center=(0,0,0))
  BF4.rotate('z',pi/3,center=(0,0,0))
  BF4.translate(slab.positions[5]+(0,0,5))
  #Put BF4 above the metal slab
  slab += BF4
  #Fix second and third layers in the slab
  mask = [atom.tag > 1 for atom in slab]
  slab.set_constraint(FixAtoms(mask=mask))
  #Setup calculator
  calc = GPAW(
    h=0.16,        #grip spacing
    kpts=(4,4,1),  #number of k-points
    txt='22.txt',  #output file
    parallel={'domain':4}, #parallelization
    xc='RPBE')     #functional
  slab.set_calculator(calc)
  qn=optimize.QuasiNewton(slab,
              trajectory='22.traj')
  qn.run(fmax=0.05)
\end{lstlisting}

The script specifies a system, a calculator (GPAW) and an algorithm for geometry optimization (QuasiNewton). The model consists of a molecular fragment (\texttt{BF4 = Atoms([Atom('B', ( 0, 0, 0)), ...)}) and an infinite gold slab (\texttt{slab = fcc111('Au', size=(2,2,3), ...}). Two layers in the gold slab are kept fixed in their bulk positions (\texttt{slab.set\_constraint()}). For the GPAW calculator we specify grid spacing (\texttt{h=0.16}) and number of $k$-points (\texttt{kpts=(4,4,1)}) at 4$\times$4$\times$1 Monkhorst-Pack sampling grid used to perform Brillouin-zone integrations. In this script we use revised Perdew--Burke--Ernzerhof (RPBE) exchange-correlation functional, but we also tested a superior vdW-DF functional to take into account the dispersive (van der Waals) interactions \cite{ivanistsev2012}. A convergence criteria for the structural optimization according to QuasiNewton algorithm is set to 0.05 eV per \r{A}ngstr\"{o}m for atomic forces (\texttt{fmax=0.05}). 



\subsection{Computational experiments performed and Results\label{results}}

In the conducted computational experiments, we used parameters similar to those shown in the script. Starting from a given geometry, the GPAW calculator was used to find an optimal geometry with the lowest energy. The obtained electron density was used to recalculate absolute electrode potential. The obtained energy values were used to estimate the interfacial free energy change via an approach introduced in Ref. \cite{rossmeisl}. Note, $\Delta G_\mathrm{int}$ was calculated using the values consistent with the forces presented in the models and neglecting contribution of entropy terms.

The capacitance was determined from the interfacial free energy dependence on the electrode potential. Ideally, $\Delta G_\mathrm{int}$ values should fit to a quadratic curve according to the equation \ref{eq:energy}: $\Delta G_\mathrm{int} = \frac12 C\Delta U^2$, where $\Delta G_\mathrm{int}$ is equal to the energy stored in an ideal capacitor which, in our case, is set up by ions and the counter charge on the gold surface. $\Delta U = (U - U_\mathrm{pzc})$, where $U$ is the electrode potential calculated from the work function ($U = W_e/e$) and $U_\mathrm{pzc}$ is the Potential of Zero Charge, taken to be equal to 5.3~V for our models \cite{ivanistsev2012}.

\begin{figure}
\centering
\includegraphics[width=0.65\textwidth]{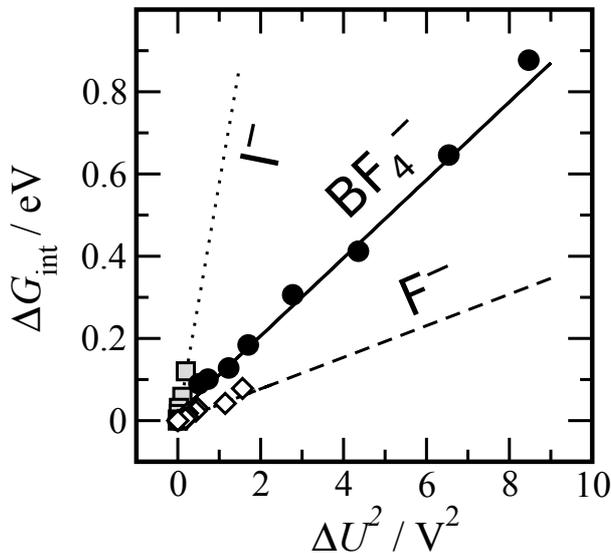}
\caption{Dependence of  interfacial free energy on electrode potential for molecular models of $\mathrm{BF}_4^-$,  $\mathrm{F}^-$ and $\mathrm{I}^-$ at Au(111) surface. The slope is proportional to the capacitance in case of an ideal parallel plate capacitor}
\label{fig:GU2}
\end{figure}

Figure \ref{fig:GU2} illustrates the relation depicted in Equation \ref{eq:energy} for the simplified interface models containing BF$_4^-$, F$^-$ and I$^-$ anions. The highest values for each of ions correspond to the formation of the ordered $\sqrt{3}\times\sqrt{3}$ adlayers (similar to the $\mathrm{BF}_4^-$ layer shown in Fig. \ref{fig:structureModel1}). The obtained values for $\mathrm{BF}_4^-$ are in a reasonable agreement with the experimental values for high frequency capacitance: 5--10 $\mu$F/cm$^2$ \cite{gnahm2010,Gnahm2011,Su2013} and for lower frequency capacitance at anodic peak 10--50 $\mu$F/cm$^2$ \cite{gnahm2010,Gnahm2011,Su2013} measured at gold surfaces.

The intefacial free energy is rising due to the repulsion between ions upon their accumulation near the surface. In the $\Delta G_\mathrm{int}$ vs. $U^2$ representation the slope at the $\Delta G_\mathrm{int}$,$\Delta U^2$-dependence corresponds to the capacitance. For a hypothetical Au(111) \textbar~F$^-$ interface, the evaluated capacitance value is lower when compared to the value for Au(111) \textbar~$\textnormal{BF}_4^-$ at the same electrode potential. That is a simple dependence on the composition of the ionic liquid. Herewith, among BF$_4^-$, F$^-$ and I$^-$ anion the iodide contributes to the highest capacitance at the interface.

Adlayers of iodide were observed in aqueous electrolytes and recently were found to form in ionic liquids \cite{Siinor2013}. As well as in the presented results for the simple model, at the laboratory conditions the formation of an adlayer is associated with the increase of capacitance. 
The difference between iodide and tetrafluoroborate ions is seen from Table \ref{tab:charges} showing the partial ionic charge evaluated according to Bader charge analysis. The redistribution of electric charge densities are explicitly treated in DFT calculations, but until now were not accounted in modern theories and classical molecular dynamics simulations \cite{dommert2012}. It is seen that iodide ions are involved in so-called partial charge transfer, when the absolute value partial ionic charge is reduced (in the case of iodide at low coverages to $-0.13\ e$) due to the interaction with the surface. Oppositely, the charge on BF$_4^-$ anion near the surface ($\approx -0.75\ e$) remains almost as high as in neat EMImBF$_4$ ionic liquid ($-0.9\ e$). At very high potentials corresponding to high coverages iodide clearly transforms into a I$^{\cdot}$ radical and BF$_4^-$ dissociates according to the net reaction: BF$_4^-$ = F$^-$ + BF$_3$, where F$^-$ has a lower partial charge and tend to react with the ionic liquid. The similar reaction mechanism was identified in recent X-ray photoelectron spectroscopy study \cite{Tonisoo2013}.

\begin{table}
\begin{center}
\begin{tabular}{ r r r r r }
\hline
Coverage			& 0.05		& 0.125   	& 0.333		& 0.375	\\
\hline
$[ \mathrm{EMIm} ] ^{\delta+}$	& 0.79		& --		& --		& --	\\
$[ \mathrm{BF}_4 ] ^{\delta-}$	& 0.85		& 0.75		& 0.69		& 0.61	\\
F$^{\delta-}$			& 0.61		& 0.61		& 0.59		& --	\\
I$^{\delta-}$			& 0.13		& 0.13		& 0.13		& 0.03	\\
\end{tabular}
\end{center}
  \caption{Partial ionic charge ($\delta\ /\ e$) obtained via Bader-analysis of the all-electron density.}
  \label{tab:charges}
\end{table}


To conclude, the conducted computational experiments shed light on the following aspects of charged interfaces in ionic liquids: 1) the capacitance values are determined by the strong repulsion between counter-ions; 2) high capacitance values are expected for ionic liquids containing iodide ions; 3) the stability of ionic liquids on the potential scale, at anode, is determined by electroreduction of anions. We note that the analysis of the results was conducted in close collaboration with the electrochemical group at the University of Tartu, in which the proposed aspects were verified in due course \cite{Siinor2013,Tonisoo2013}.

\section{Evaluation of the migration process and the D2CM tool}

Once the first version of D2CM tool was completed (and before the migration of the larger case study) we initiated a pilot project as a cooperation between computer scientists working with the D2CM tool and the scientists working in the electrochemistry field. Our primary goal was to validate the migration and deployment functionalities of the D2CM tool by verifying that the migrated systems are working as intended and that the results of the experiments are correct. We also aimed to make sure that the tool is fully usable by non-computer scientists. 

We asked scientists working in the electrochemistry field to design a number of computational experiments that could be executed on the cloud using our tool. At the time of project initiation, corresponded to the state of the art in the field of electrochemistry. Table~\ref{tab:GPAW_EC2} summarizes a small subset of these experiments that have been conducted by us since 2010. Rows 1 to 4 describe the results of running a single experiment on varying cluster sizes and rows 5 to 8 describe randomly chosen computations in different scales. 

\begin{table}
\begin{center}
\begin{tabular}{ r r r r r r r r r  }
\hline
Experiment	& EC2 Instance type 	& Nodes   	& EC2		& RAM 		& Time  	& Cost  \\
ID		&		& (Cores) 	& Units		& (GB) 		& (Hrs) 	& (USD) \\\hline
dod1		& m1.medium	& 1 (1)		& 2		& 3.75		& 3.10		& 0.48\\
dod2		& m1.medium	& 2 (2)		& 4		& 7.5		& 1.60		& 0.48\\
dod4		& m1.medium	& 4 (4)		& 8		& 15		& 0.87		& 0.48\\
dod8		& m1.medium	& 8 (8)		& 16		& 30		& 0.42		& 0.96\\
22 		& m1.xlarge	& 1 (4)		& 8		& 15.0		& 4.3		& 3.80 	\\
22vdw 		& m1.xlarge	& 2 (8)		& 16		& 30.0		& 1.0		& 1.52 	\\
32 		& m1.xlarge	& 2 (8)		& 16		& 30.0		& 45.7		& 69.92 \\
34 		& m1.xlarge	& 8 (32)	& 64		& 120.0		& 46.3		& 285.76\\
\end{tabular}
\end{center}
  \caption{A sample of GPAW experiments performed on EC2 using D2CM}
  \label{tab:GPAW_EC2}
\end{table}

It must be noted, that many of the current D2CM features were actually envisioned during this initial pilot project which involved intensive cooperation of the electrochemists and the D2CM developers. Once the pilot project was completed and the additional functionalities requested by the scientists were introduced, the consecutive experiments were performed by the electrochemists themselves without taking the help of computer scientists. While some of the results of these studies are already addressed in section \ref{casestudy}, in the context of the case study, the more detailed results will be submitted to a domain specific journal \cite{ivanistsev2012}.

One example of a D2CM feature that was introduced as result of cooperation with electrochemists is performance monitoring. Because electrochemical experiments (and especially the more complex ones) can run for extended periods of time the scientist needed a way to verify that the migrated experiment is working as intended and that the chosen deployment configuration was suitable. Thus we introduced real-time performance monitoring to the D2CM tool which allows the users to keep track of how well the computing resources are utilized. 

\begin{figure}
\centering
\includegraphics[width=0.70\textwidth]{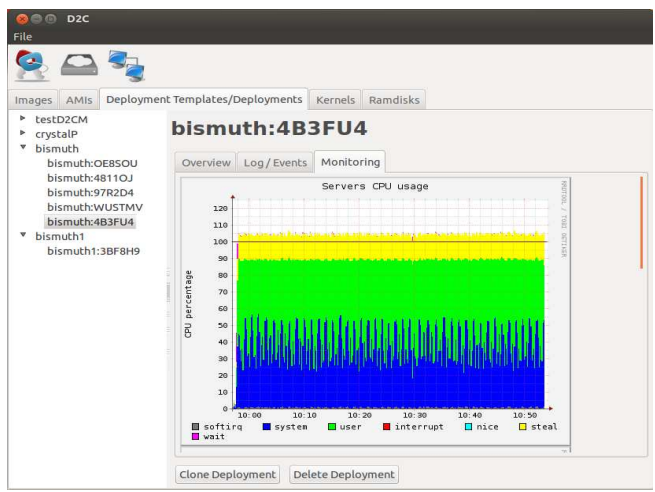}
\caption{CPU profile of a computational experiment}
\label{fig:CPUProfile}
\end{figure}

\begin{figure}
\centering
\includegraphics[width=0.70\textwidth]{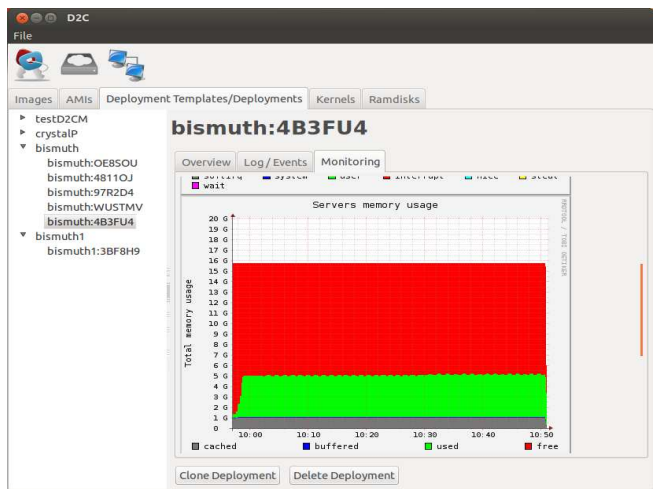}
\caption{Memory profile of a computational experiment}
\label{fig:MemProfile}
\end{figure}

Figures \ref{fig:CPUProfile} and \ref{fig:MemProfile} show the CPU and memory profile of an computational experiment running on 4 m1.medium EC2 instances. D2CM uses \texttt{collectd} \cite{collectd}, an industry standard in distributed system monitoring, for extracting the performance parameters from the instances.  The tool records CPU usage by type (user, system, idle and steal), memory usage, network I/O and system load for the whole deployed cluster. This feature can be turned off to save the system resources.

Figure \ref{fig:CPUProfile} shows the average CPU usage of the hosts and the figure \ref{fig:MemProfile} depicts total memory usage across the hosts for one of the executed experiments. From these diagrams we could see that the CPU was utilized to the full capacity (with a significant amount of CPU steal) while memory was significantly underutilized. Thus it is clear that this experiment would significantly gain from a more capable CPU and that does not need as much memory. So the subsequent computational experiments ran on High-CPU instance family of EC2, which have more powerful CPU and less memory than the standard instance family member m1.medium used in the previous computational experiment. Thus, the functionality that was initially introduced for keeping track of long running experiments also significantly helps in optimizing the deployed compute clusters, computational experiments' set-up and the costs of the experiments. 

In terms of costs of cloud computing resources, the usage of Amazon EC2 is still more justifiable for enterprise needs, while the usage of supercomputers or grids instead of clouds is still more suitable for academic users. For enterprise sector the cost of CPU-time at a supercomputer and EC2 is comparable. For instance, price per core hour at a regional supercomputer in UK is \$0.16 for a Dual Intel Xeon X5650 2.66 GHz CPU \cite{archie}. Price per core hour at EC2 is \$0.16 for Intel Xeon X5570 (4-core Nehalem architecture; Cluster Compute Quadruple Extra Large Instance) \cite{amazon:price}. On the other hand, for typical scientific computing environments, there is no long queue to access shared infrastructure with D2CM. Thus, we may expect that in the near future, the D2CM tool will be used more often for academic purposes. Procuring and provisioning cost-effective computation cycles is possible via D2CM through optimizing the compute clusters; improving scalability by adjusting GPAW set-up; and utilizing Amazon EC2 Spot Instances. Overall, D2CM provides a good basis for effective high-performance calculations in electrochemistry as well as in the other scientific fields.




\subsection{Scalability of computational experiments on cloud}

To validate that the D2CM tool is able to utilize the distributed cloud resources in an efficient manner and to illustrate its usability for running distributed scientific experiment we also performed scalability tests. For measuring the scalability of the deployed system we once again used GPAW based experiments because GPAW supports a number of different means for defining the parallelism. It relies on the MPI for parallelization and can be rebuilt with optimized libraries for parallel computing such as Scalable Linear Algebra PACKage (ScaLAPACK) \cite{scalapack}.

We chose a medium-size computational experiment and measured its runtime when deployed to EC2 cloud on clusters from 1 to 16 cloud m1.medium instances with ``moderate'' network performance (as assessed by AWS), 3.75 GB of memory and 2 EC2 Compute
Units (1 virtual core). EC2 computing unit represents the capacity of allocated processing power and 1 EC2 unit is considered to be equivalent to an early-2006 1.7 GHz Xeon processor. 

\begin{figure}
\centering
\includegraphics[width=0.70\textwidth]{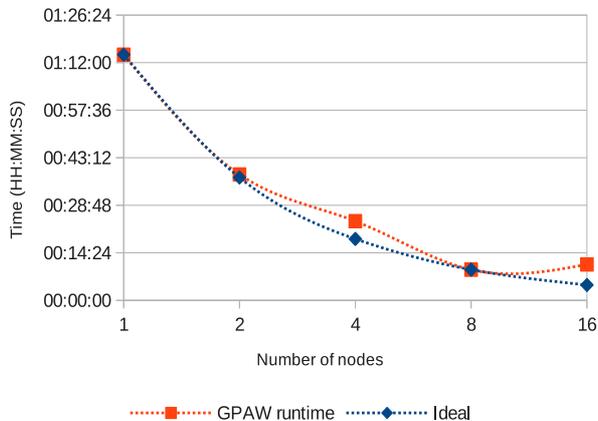}
\caption{Runtime diagram of a GPAW scalability test}
\label{fig:scale1}
\end{figure}

\begin{figure}
\centering
\includegraphics[width=0.70\textwidth]{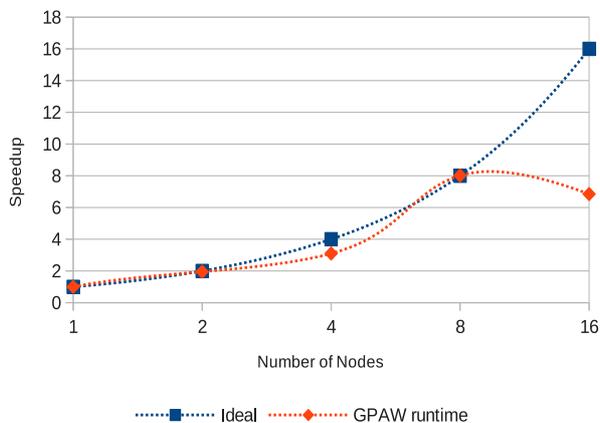}
\caption{Speedup diagram of a GPAW scalability test}
\label{fig:scale}
\end{figure}

We measured the runtime of 20 optimization steps. The results are illustrated on figures \ref{fig:scale1} and \ref{fig:scale}. Figure  \ref{fig:scale1} shows a runtime of a GPAW experiment when deployed on a varying cluster size from 1 to 16 nodes. The ideal curve is calculated assuming perfect scaling, which is possible with a high degree of software parallelization and excellent quality of the network. Figure \ref{fig:scale} provides the parallel speedup, which shows how much gain do we get from introducing additional nodes. While this experiment does not gain any benefit from using 16 nodes over 8 nodes (showing the problem size might be too small to be deployed on 16 nodes), it does demonstrate that a virtual cloud cluster with moderate network performance provides good parallelization performance up to 8 nodes.

Some parallelization techniques are common among a number of program packages for performing electronic structure calculations, like GPAW, WIEN2k \cite{wien2k},  Quantum Espresso \cite{QE2009}, FEFF9 \cite{FEFF9} as well as many others, for example, those sharing Libxc-library and listed in reference \cite{libxc}. For this reason it might be expected that similar scalability, as seen in tests
with GPAW, can be expected for the listed codes. Scalability tests made by Jorissen, Vila and Rehr for WIEN2k and FEFF9 codes complement our observation that a virtual cloud cluster provides parallelization performance as a physical cluster with very low latency and high bandwidth of the cluster interconnects \cite{Jorissen2012}. Moreover, Jorissen et al. indicated a case when the network performance of regular EC2 instances becomes an issue, as in our test \cite{Jorissen2012}. The problem arises due to the MPI intensive communication needed for large dense matrix diagonalizations employing the ScaLAPACK library \cite{scalapack}. On the cloud side it can be resolved by utilization of HPC EC2 instances placed into a non-blocking 10 Gigabit ethernet network. In general the problem persists for systems with a large number of electrons. Therefore, we conclude that virtual cloud clusters provide a good speedup performance suitable for performing small to medium computational electrochemistry experiments. 

\subsection{Usability analysis of the D2CM tool and the cloud migration}

Apart from the technical evaluation of the D2CM tool based migration procedure, we are also interested in its usability. In general, for research groups having a permanent heavy load of computational tasks, a private cluster is more justified than a cloud from the cost point of view. However, for small and medium research groups with lower intensity in computing and only basic expertise in HPC, it makes better sense to pay for resources and support on demand, from the electrochemists' perspective. D2CM tool provides an easy way for this cloud migration and it simplified running the actual computational experiments for the scientists, by allowing them to change the deployment configurations and computational experiment parameters on the fly. Feedback from the scientists who have been working with the tool show that they are quite satisfied with the tool and the migration procedure. 


In addition to having a tool for cloud migration, to take full advantage of the cloud a scientist may also have to consider the following aspects. 
Firstly, the cloud solution can be used solely or to augment the available facility (cluster, GRID or HPC), e.g. when the facility is overloaded. Secondly, a large number of universities and companies have neither an in-house cluster nor the resources to build and maintain one. Access to the local GRID or HPC facilities might be limited due to a queuing system, specific restrictions, overload, etc. In such circumstances cloud provides access to a much larger number of CPUs that would usually be available to solve problems that are time-critical. Finally, one can use the cloud to run software that requires a different environment than the one existing on the available facility. With the D2CM tool it is easy and quick to build the cloud environment and start using it, as if it was a desktop environment. Moreover, the tool also gives the extra flexibility at keeping track of the running costs and making them as low as possible.

The most obvious competitive edge of the cloud solution is its ability to explore specific research tasks in a timely, scalable manner and helps scientist to make better decisions quickly. Let us stress that a cloud can serve to augment existing clusters when the load on the cluster is too high. This way, the pay-per-use model can positively change the dynamics of computations in a research group. Even despite the higher price for the CPU time in common clouds, it can be very useful when time-to-result is a priority.

However, with the cloud migration, the type of computation and the software should be deliberately considered. On the same number of nodes there might be a large number of simultaneous small calculations or a single massively parallel calculation. From the practical point of view, the cloud computing is more efficient in the first case, but might struggle to fulfill a massively parallel job due to low interconnectivity between the nodes~\cite{srirama2011scalability}. Next, precise estimation of time and the cost is only possible in some types of calculations, for instance classical molecular dynamics simulations \cite{Ebejer2013}. It is a challenging, yet solvable task to estimate the costs for example for quantum chemical calculations, like in the presented computational electrochemical study. 

Considering all these advantages and the successful migration of the electrochemistry case using the D2CM tool, currently discussions are underway to make the tool available to the GPAW community. The tool will soon be put on the GPAW site~\cite{GPAW}. The tool is already released with a GPL License~\cite{Srirama:MCLab.D2CM}.

\section{Related Work}

D2CM tool provides three core features: migrating desktop virtual machines to IaaS platforms, specifying deployment configurations and life-cycle management of deployed system. While these features are present to some extent in other tools, D2CM is the only tool, as far as we know, that supports the whole process from migration to deployment and monitoring in one package and is specifically aimed for performing distributed scientific experiments in the cloud.

The main reason for developing the D2CM tool is that the migration of desktop images to the cloud has not been a widely supported feature by the cloud industry in the past. However, several tools have been recently published that do aim to provide this service. Amazon has developed VM Import \cite{vmimport} tool for migration of VMware ESX VMDK, Citrix Xen VHD and Microsoft Hyper-V VHD images to Amazon EC2 cloud. In addition to only supporting images from select proprietary systems, the VM Import tool currently supports only images running Microsoft Windows Server. Amazon has stated their commitment to add more supported platforms, but a concrete roadmap has not been published yet. 

Racemi Cloud Path \cite{racemi} is a solution that provides tools for migrating existing server environments, that run Red Hat Enterprise Linux or CentOS operating systems, to Amazon EC2, GoGrid, Rackspace, and Terremark public clouds. It automates the otherwise manual process of creating a virtual machine (VM) from an existing physical machine configuration and migrating VM to the cloud. Racemi Cloud Path is not a free service and while it simplifies the migration process, it lacks the support for custom deployment configurations for the migrated virtual machines. Also, setting up already migrated environments has to be done manually, which can be a huge amount of work depending on the complexity of the deployment, especially when the number of different interacting roles and instances is high.   

CloudSwitch \cite{cloudswitch} provides migration and deployment of virtual machines to the cloud and allows to integrate migrated servers with an existing system that runs locally or in other clouds. 
It is designed for setting up (or extending) longer running static server deployments. Defining and configuring different roles and instance configurations must be repeated each time a new deployment is executed. 
In comparison, once a deployment template (consisting of any number of roles, life-cycle actions, and instances) is defined in D2CM, the user can instantiate a new deployment, change its configuration parameters and then launch it, all in only a few seconds. 

RightScale offers means of managing systems via deployment definitions, which make use of RightScale ServerTemplates \cite{rightscale:servertemplate}. ServerTemplate defines the machine image and scripts executed during the startup and shutdown phases of a machine. Amazon also provides EC2 system templating through the CloudFormation \cite{amazon:cloudformation} tool. CloudFormation allows users to describe the images and machine types that comprise the logical roles of a system. While CloudFormation allows for reliable and somewhat flexible deployment of a system, any life-cycle management must be explicitly programmed into the VM images themselves. On Linux, this is supported by Amazon through the CloudInit software package. This conflicts with our approach in D2CM, where we seek to minimize the modification of the source images by the user, as our targeted users are pure scientists who are not proficient in managing server environments. 

Similarly, other cloud frameworks that provide capabilities for automatic provisioning, deployment and monitoring of cloud systems are Cloudify \cite{cloudify}, Puppet \cite{puppet} and Chef \cite{chef}. However, while they are good for deploying more static systems on the cloud and might be suitable for setting up more permanent scientific clusters, they are not suitable for running scientific experiments with varied duration and they also do not provide means to define the life-cycle of such experiments for their automatic execution.    

VIRTU framework \cite{DuroSLPM10} is a framework for configuration, deployment, and management of the life-cycle of distributed virtual environments. It provides users the ability to manage and configure application stacks on such environments by assembling building blocks consisting of operating system and application together with installation and configuration files. This framework is interesting for us because it deals with the deployment of distributed environments that large-scale scientific computing experiments most often require, with the main difference that it is not designed for Cloud, but rather for deploying virtual machines and applications on existing hardware.  


Cloud Modelling Framework (CloudMF) \cite{Ferry:2013:MMS:2513534.2513542} is a model-based framework being developed at SINTEF (http://www.sintef.no) for modeling the provisioning of cloud resources for complex software systems and their deployment and control in multi-cloud platforms. CloudMF provides a tool-supported domain-specific modeling language CloudML to model the provisioning and deployment of multi-cloud systems and a models@run-time environment for enacting the provisioning, deployment, and adaptation of these systems. The deployment of the underlying modeling language CloudML has been partially supported by three European FP7 projects (REMICS, MODAClouds and PaaSage), where we were also involved, which all try to leverage its potential for simplifying the deployment and control of cloud applications. 

While the modeling approach taken by CloudML can be very useful in the software engineering field and can really benefit the adoption of cloud platforms for mode complex and especially legacy systems, the aim of the D2CM tool is to simplify the migration and deployment of scientific experiments for scientists who often have no prior knowledge of software engineering and modeling techniques. Still, we are investigating whether we can utilize CloudMF internally to streamline the deployment of scientific experiments in multi-cloud platforms while hiding all the technical details from the users. Our future research in this domain will try to address this issue. 




\section{Conclusions and future work}

This paper discusses the direct migration of scientific computing applications to the cloud using the D2CM tool. The main goal is to migrate the complete software environment, in which the scientists have set up their computational experiments, directly to the cloud.  The tool also facilitates and simplifies the execution of the experiments in the migrated system for non-computer scientists. The design and development of the D2CM tool is explained in detail. In the paper we also present an electochemical case study targeted at designing modern supercapacitors. We discuss the role of D2CM tool in achieving domain specific results, the experiments performed, as well as the cloud scalability.  



From the analysis of the case study, we observed that D2CM tool not only helps in the migration process but also simplifies deploying MPI applications in the cloud, helps to optimize the calculations, compute clusters and the costs of the experiments. Feedback from the users has shown that the D2CM tool also simplifies the actual experiment execution process by allowing the user to change the deployment configurations and experiment parameters on the fly. The tool and the simplified migration procedure can help the domain scientists avoid long queues at the grid resources and provide them a new dimension, cost-to-value of the computational experiments.

Our current work involves extracting scaling parameters for the calculations and using them to predict the cost of future experiments. The pricing schemes in public clouds are often very complex to measure. For example, Amazon EC2 does not only measure instance (virtual machine) type and running time but also network bandwidth, static IP idle time, size of persistent data stored in the cloud, etc. This is very important when running scientific computing experiments in the cloud as the size and parameters of the calculations are expected to change frequently and thus may require different amounts of computing resources every time. For this reason, it is important to estimate the computing resources needed for new experiments based on the input size and parameters used, and to estimate the cost of the deployment based on the approximated calculation time. However, the scale up parameters can only be estimated on case by case basis. We are currently in the process of developing a feasible solution using the linear programming model.
Apart from this, as already mentioned in the related work section, we are also investigating how to integrate the multi-cloud deployment of CloudMF into D2CM.

\section*{Acknowledgment}
This research is supported by European Commission via the REMICS project (FP7-257793), Estonian Science Foundation grants ETF9287 and ETF8357, Estonian IT Academy, European Regional Development Fund through the EXCS and Target Funding theme SF0180008s12. The authors would also like to thank Chris Willmore and Ulrich Norbisrath who are involved with the initial development of the D2CM tool.

\bibliographystyle{plain}
\bibliography{bibliography}

\begin{thebibliography}{10}

\bibitem{chef}
Chef.
\newblock \url{http://www.getchef.com/chef/}, November 2015.

\bibitem{libguestfs}
Libguestfs, a toolkit for modifying virtual machine disk images.
\newblock \url{http://libguestfs.org/}, November 2015.

\bibitem{libvirt}
Libvirt virtualization toolkit.
\newblock \url{http://libvirt.org/}, November 2015.

\bibitem{nobel1998}
Press release: The 1998 nobel prize in chemistry.
\newblock
  \url{http://www.nobelprize.org/nobel_prizes/chemistry/laureates/1998/press.html},
  November 2015.

\bibitem{nobel2013}
Press release: The 2013 nobel prize in chemistry.
\newblock
  \url{http://www.nobelprize.org/nobel_prizes/chemistry/laureates/2013/press.html},
  November 2015.

\bibitem{amazon:cloudformation}
{Amazon Inc.}
\newblock Amazon cloud formation.
\newblock \url{http://aws.amazon.com/cloudformation/}, November 2015.

\bibitem{amazon:price}
{Amazon Inc.}
\newblock Amazon cloud pricing.
\newblock \url{http://aws.amazon.com/ec2/pricing/}, November 2015.

\bibitem{Amazon:EC2}
{Amazon Inc.}
\newblock Amazon elastic compute cloud (amazon ec2).
\newblock \url{http://aws.amazon.com/ec2/}, November 2015.

\bibitem{amazon:s3}
{Amazon Inc.}
\newblock Amazon simple storage service.
\newblock \url{http://aws.amazon.com/s3/}, November 2015.

\bibitem{vmimport}
{Amazon Inc.}
\newblock Amazon virtual machine import and export service.
\newblock \url{http://aws.amazon.com/ec2/vmimport/}, November 2015.

\bibitem{amazonkernels}
{Amazon Inc.}
\newblock Enabling user provided kernels in amazon ec2.
\newblock \url{http://aws.amazon.com/articles/3967}, November 2015.

\bibitem{Anderson2012}
A.~B. Anderson.
\newblock Insights into electrocatalysis.
\newblock {\em Phys. Chem. Chem. Phys.}, 14(4):1330--1338, 2012.

\bibitem{ase}
S.~R. Bahn and K.~W. Jacobsen.
\newblock An object-oriented scripting interface to a legacy electronic
  structure code.
\newblock {\em Comput. Sci. Eng.}, 4(3):56--66, 2002.

\bibitem{Barham:2003:XAV:1165389.945462}
P.~Barham, B.~Dragovic, K.~Fraser, S.~Hand, T.~Harris, A.~Ho, R.~Neugebauer,
  I.~Pratt, and A.~Warfield.
\newblock Xen and the art of virtualization.
\newblock {\em SIGOPS Oper. Syst. Rev.}, 37(5):164--177, 2003.

\bibitem{scalapack}
L.~S. Blackford, J.~Choi, A.~Cleary, E.~D'Azevedo, J.~Demmel, I.~Dhillon,
  J.~Dongarra, S.~Hammarling, G.~Henry, A.~Petitet, K.~Stanley, D.~Walker, and
  R.~C. Whaley.
\newblock {\em {ScaLAPACK} Users' Guide}.
\newblock Society for Industrial and Applied Mathematics, Philadelphia, PA,
  1997.

\bibitem{wien2k}
P.~Blaha, K.~Schwarz, G.~Madsen, D.~Kvasnicka, and J.~Luitz.
\newblock {\em WIEN2k, An Augmented Plane Wave + Local Orbitals Program for
  Calculating Crystal Properties}.
\newblock Karlheinz Schwarz, Techn. Universit\"at Wien, Austria, 2001.

\bibitem{mapscale}
C.~Bunch, B.~Drawert, and M.~Norman.
\newblock {MapScale: A Cloud Environment for Scientific Computing}.
\newblock Technical report, University of California, Computer Science
  Department, 2009.

\bibitem{SPE:SPE995}
R.~N. Calheiros, R.~Ranjan, A.~Beloglazov, C.~A.~F. {De Rose}, and R.~Buyya.
\newblock Cloudsim: a toolkit for modeling and simulation of cloud computing
  environments and evaluation of resource provisioning algorithms.
\newblock {\em oftware Pract. Exper.}, 41(1):23--50, 2011.

\bibitem{Calle-vallejo2012}
F.~Calle-Vallejo and M.~{T.M.} Koper.
\newblock First-principles computational electrochemistry: Achievements and
  challenges.
\newblock {\em Electrochim. Acta}, 84:3--11, 2012.

\bibitem{GPAW}
{Center for Atomic-scale Materials Design}.
\newblock {GPAW}.
\newblock \url{http://wiki.fysik.dtu.dk/gpaw/}, November 2015.

\bibitem{cloudswitch}
{CloudSwitch, Inc.}
\newblock Cloudswitch.
\newblock \url{http://www.cloudswitch.com/}, November 2015.

\bibitem{Conway1999}
B.~E Conway.
\newblock {\em Electrochemical supercapacitors: scientific fundamentals and
  technological applications}.
\newblock Springer, 1999.

\bibitem{dommert2012}
F.~Dommert, K.~Wendler, R.~Berger, {S. L.} Delle, and C.~Holm.
\newblock Force fields for studying the structure and dynamics of ionic
  liquids: A critical review of recent developments.
\newblock {\em {ChemPhysChem}}, 13(7):1625--1637, 2012.

\bibitem{DuroSLPM10}
N.~Duro, R.~Santos, J.~Louren\c{c}o, H.~Paulino, and J.~Martins.
\newblock Open virtualization framework for testing ground systems.
\newblock In {\em PDATAD}, pages 67--73. ACM, 2010.

\bibitem{Ebejer2013}
J.-P. Ebejera, S.~Fullea, G.~M. Morrisa, and P.~W. Finn.
\newblock The emerging role of cloud computing in molecular modelling.
\newblock {\em J. Mol. Graph. Model.}, 44:177 -- 187, 2013.

\bibitem{GPAW2}
J.~Enkovaara, C.~Rostgaard, J.J. Mortensen, M.~Kuisma, and et~al.
\newblock Electronic structure calculations with {GPAW:} a real-space
  implementation of the projector augmented-wave method.
\newblock {\em J. Phys. Condens. Mat.}, 22(25):253202, 2010.

\bibitem{Euca:site}
{Eucalyptus Systems Inc.}
\newblock Eucalyptus.
\newblock \url{http://www.eucalyptus.com}, November 2015.

\bibitem{Fedorov2014}
M.~V. Fedorov and A.~A. Kornyshev.
\newblock Ionic liquids at electrified interfaces.
\newblock {\em Chem. Rev.}, 114(5):2978--3036, 2014.

\bibitem{Ferry:2013:MMS:2513534.2513542}
N.~Ferry, F.~Chauvel, A.~Rossini, B.~Morin, and A.~Solberg.
\newblock Managing multi-cloud systems with cloudmf.
\newblock In {\em Proceedings of the Second Nordic Symposium on Cloud Computing
  \#38; Internet Technologies}, NordiCloud '13, pages 38--45, New York, NY,
  USA, 2013. ACM.

\bibitem{FutureGrid.13}
{FutureGrid}.
\newblock {Using IaaS Clouds on FutureGrid}.
\newblock {\url{https://portal.futuregrid.org/using/clouds}}, November 2015.

\bibitem{QE2009}
P.~Giannozzi, S.~Baroni, and et~al.
\newblock Quantum espresso: a modular and open-source software project for
  quantum simulations of materials.
\newblock {\em J. Phys. Condens. Mat.}, 21(39):395502 (19pp), 2009.

\bibitem{cloudify}
{GigaSpaces Technologies}.
\newblock Cloudify.
\newblock
  \url{http://www.gigaspaces.com/cloudify-cloud-orchestration/overview},
  November 2015.

\bibitem{Gnahm2011}
M.~Gnahm, C.~M{\"u}ller, R.~R{\'e}p{\'a}nszki, T.~Pajkossy, and {D. M.} Kolb.
\newblock The interface between {Au}(100) and
  1-butyl-3-methyl-imidazolium-hexafluorophosphate.
\newblock {\em Phys. Chem. Chem. Phys.}, 13:11627--11633, 2011.

\bibitem{gnahm2010}
M.~Gnahm, T.~Pajkossy, and {D.M.} Kolb.
\newblock The interface between {Au(111)} and an ionic liquid.
\newblock {\em Electrochim. Acta}, 55(21):6212--6217, 2010.

\bibitem{HelixNebula.13}
{Helix Nebula}.
\newblock {The Science Cloud}.
\newblock {\url{http://www.helix-nebula.eu/}}, November 2013.

\bibitem{archie}
{High Performance Computing for the West of Scotland}.
\newblock The {ARCHIE-WeSt} service for industry users, November 2015.

\bibitem{Ivanistsev2015}
V.~Ivani{\v s}t{\v s}ev, K.~Kirchner, T.~Kirchner, and M.~V. Fedorov.
\newblock Restructuring of the electrical double layer in ionic liquids upon
  charging.
\newblock {\em J. Phys. Condens. Mat.}, 27(10):102101, 2015.

\bibitem{ivanistsev2012}
V.~Ivani{\v s}t{\v s}ev, V.~Tripkovi{\'c}, and J.~Rossmeisl.
\newblock Density functional theory study of a charged {Au}(111) {EMImBF}$_4$
  (ionic liquid) interface.
\newblock in prep., 2015.

\bibitem{Jorissen2012}
K.~Jorissen, W.~Johnson, {F.D.} Vila, and {J.J.} Rehr.
\newblock High-performance computing without commitment: {SC2IT:} a cloud
  computing interface that makes computational science available to
  non-specialists.
\newblock In {\em 2012 {IEEE} 8th International Conference on E-Science
  (e-Science)}, pages 1--6, 2012.

\bibitem{kurig2011}
H.~Kurig, M.~Vestli, A.~J{\"a}nes, and E.~Lust.
\newblock Electrical double layer capacitors based on two
  {1-Ethyl-3-Methylimidazolium} ionic liquids with different anions.
\newblock {\em Electrochem. Solid State Lett.}, 14(8):A120--A122, 2011.

\bibitem{lewandowski2004}
A.~Lewandowski and M.~Galinski.
\newblock Carbon-ionic liquid double-layer capacitors.
\newblock {\em J. Phys. Chem. Sol.}, 65(2-3):281--286, 2004.

\bibitem{DBLP:conf/synasc/LiG10}
Q.~Li and Y.~Guo.
\newblock Optimization of resource scheduling in cloud computing.
\newblock In {\em Symbolic and Numeric Algorithms for Scientific Computing
  (SYNASC), 2010 12th Int. Symp. on}, pages 315 --320, 2010.

\bibitem{lust2003}
E.~Lust, G.~Nurk, A.~J{\"a}nes, M.~Arulepp, P.~Nigu, P.~M{\"o}ller, S.~Kallip,
  and V.~Sammelselg.
\newblock Electrochemical properties of nanoporous carbon electrodes in various
  nonaqueous electrolytes.
\newblock {\em J. Solid State Electrochem.}, 7(2):91--105, 2003.

\bibitem{ArmbrustETAL:AboveCloud.09}
{M. Armbrust et al.}
\newblock Above the clouds: A berkeley view of cloud computing.
\newblock Technical report, University of California, 2009.

\bibitem{libxc}
M.~{A. L.} Marques, M.~{J. T.} Oliveira, and T.~Burnus.
\newblock Libxc: A library of exchange and correlation functionals for density
  functional theory.
\newblock {\em Comp. Phys. Commun.}, 183(10):2272--2281, 2012.

\bibitem{collectd}
Florian octo Forster.
\newblock Collectd, a daemon for collecting system performance statistics,
  November 2015.

\bibitem{pan2006}
{G.-B.} Pan and W.~Freyland.
\newblock {2D} phase transition of {PF}$_6$ adlayers at the electrified ionic
  {liquid/Au(111)} interface.
\newblock {\em Chem. Phys. Lett.}, 427(1-3):96--100, 2006.

\bibitem{puppet}
{Puppet Labs}.
\newblock Puppet.
\newblock \url{http://puppetlabs.com/solutions/cloud-management}, November
  2015.

\bibitem{racemi}
{Racemi}.
\newblock Racemi cloud path.
\newblock \url{http://www.racemi.com/cloud-path}, November 2015.

\bibitem{FEFF9}
J.~J. Rehr, J.~J. Kas, F.~D. Vila, M.~P. Prange, and K.~Jorissen.
\newblock Parameter-free calculations of x-ray spectra with {FEFF9}.
\newblock {\em Phys. Chem. Chem. Phys.}, 12(21):5503--5513, 2010.

\bibitem{rightscale:servertemplate}
{RightScale Inc.}
\newblock Rightscale servertemplate.
\newblock \url{http://www.rightscale.com/products/configuration-framework.php},
  November 2015.

\bibitem{rossmeisl}
J.~Rossmeisl, E.~Sk\'ulason, M.~E Bj\"orketun, V.~Tripkovi\'c, and J.~K.
  N{\o}rskov.
\newblock Modeling the electrified solid-liquid interface.
\newblock {\em Chem. Phys. Lett.}, 466(1-3):68--71, 2008.

\bibitem{Siinor2013}
L.~Siinor, C.~Siimenson, K.~Lust, and E.~Lust.
\newblock Mixture of 1-ethyl-3-methylimidazolium tetrafluoroborate and
  1-ethyl-3-methylimidazolium iodide: A new potential high capacitance
  electrolyte for {EDLCs}.
\newblock {\em Electrochem. Commun.}, 35:5--7, 2013.

\bibitem{tartutech}
{Skeleton Technologies O\"{U}}.
\newblock About the {SkeletonC} technology.
\newblock \url{http://www.skeletontech.com/}, November 2015.

\bibitem{Srirama:MCLab.D2CM}
S.~N. Srirama.
\newblock Desktop to cloud migration.
\newblock \url{http://mc.cs.ut.ee/mcsite/projects/desktop-to-cloud-migration},
  November 2015.

\bibitem{SriramaETAL:SciCloud.10}
S.~N. Srirama, O.~Batrashev, and E.~Vainikko.
\newblock {SciCloud: Scientific Computing on the Cloud}.
\newblock In {\em The 10th IEEE/ACM Int.l Symposium on Cluster, Cloud and Grid
  Computing(CCGrid 2010)}, page 579, 2010.

\bibitem{srirama2013direct}
S.~N. Srirama, V.~V.~Ivani{\v s}t{\v s}ev, P.~Jakovits, and C.~Willmore.
\newblock Direct migration of scientific computing experiments to the cloud.
\newblock In {\em High Performance Computing and Simulation (HPCS), 2013
  International Conference on}, pages 27--34. IEEE, 2013.

\bibitem{srirama2011scalability}
S.N. Srirama, O.~Batrashev, P.~Jakovits, and E.~Vainikko.
\newblock Scalability of parallel scientific applications on the cloud.
\newblock {\em Scientific Programming}, 19(2):91--105, 2011.

\bibitem{Su2013}
Y.~Su, J.~Yan, M.~Li, M.~Zhang, and B.~Mao.
\newblock Electric double layer of au({100)/Imidazolium-Based} ionic liquids
  interface: Effect of cation size.
\newblock {\em J. Phys. Chem. C}, 117(1):205--212, 2013.

\bibitem{Tonisoo2013}
A.~T{\~o}nisoo, J.~Kruusma, R.~P{\"a}rna, A.~Kikas, M.~Hirsim{\"a}ki,
  E.~N{\~o}mmiste, and E.~Lust.
\newblock In situ {XPS} studies of electrochemically negatively polarized
  molybdenum carbide derived carbon double layer capacitor electrode.
\newblock {\em J. Electrochem. Soc.}, 160(8):A1084--A1093, 2013.

\bibitem{Ivanistsev2014}
V.~V.~Ivani{\v s}t{\v s}ev, S.~O'Connor, and M.~V. Fedorov.
\newblock Poly(a)morphic portrait of the electrical double layer in ionic
  liquids.
\newblock {\em Electrochem. Commun.}, 48:61--64, 2014.

\bibitem{buyya.2009}
C.~Vecchiola, S.~Pandey, and R.~Buyya.
\newblock High-performance cloud computing: A view of scientific applications.
\newblock In {\em Pervasive Systems, Algorithms, and Networks (ISPAN), 2009
  10th International Symposium on}, pages 4--16, 2009.

\end{thebibliography}

\end{document}